\documentclass[aps,prd,twocolumn,showpacs,preprintnumbers,amsmath,amssymb,floatfix,nofootinbib]{revtex4}
\usepackage{graphicx}

\usepackage{color}
\usepackage{colordvi}

\newcommand{\beqn}{\begin{eqnarray}}
\newcommand{\eeqn}{\end{eqnarray}}
\newcommand{\be}{\begin{equation}}
\newcommand{\ee}{\end{equation}}

\newcommand{\ba}{\begin{array}{c}}
\newcommand{\bat}{\begin{array}{cc}}
\newcommand{\ea}{\end{array}}

\newcommand{\bi}{\begin{itemize}}
\newcommand{\ei}{\end{itemize}}

\newcommand{\ket}{\,\rangle}
\newcommand{\bra}{\langle \,}

\newcommand{\Frac}[2]{\frac{\displaystyle #1}{\displaystyle #2}}

\newcommand{\cO}{{\cal O}}

\newcommand{\mF}{\mathcal{F}}

\newcommand{\mL}{\mathcal{L}}

\newcommand{\lsim}{\stackrel{<}{_\sim}}

\newcommand{\Int}{\displaystyle{\int}}

\newcommand{\good}{\scriptsize\mbox{`good'}}
\newcommand{\bad}{\scriptsize\mbox{`bad'}}

\begin{document}

\preprint{IFIC 09/11}
\preprint{SFB-CPP 09/27}
\preprint{PITHA 09/09}


\title{Resonance saturation of the chiral couplings at NLO in 1/$N_C$}

\author{Ignasi Rosell${}^{1,2}$}
\author{Pedro Ruiz-Femen\'\i a${}^3$}
\author{Juan Jos\'e Sanz-Cillero${}^4$}

\affiliation{${}^1$  Departamento de Ciencias F\'\i sicas, Matem\'aticas y de la Computaci\'on, Universidad CEU Cardenal Herrera, c/ Sant Bartomeu 55, E-46115 Alfara del Patriarca, Val\`encia, Spain.
}
\affiliation{${}^2$ 
IFIC, Universitat de Val\`encia -- CSIC, Apt. Correus 22085, E-46071
Val\`encia, Spain}
\affiliation{${}^3$  Institut f\"ur Theoretische Physik E, RWTH Aachen University, D-52056 Aachen, Germany
}
\affiliation{${}^4$  Grup de F\'\i sica Te\`orica and IFAE, Universitat Aut\`onoma de Barcelona, E-08193 Barcelona, Spain
}

\begin{abstract}
The precision obtainable in phenomenological applications of Chiral Perturbation Theory is currently limited
by our lack of knowledge on the low-energy constants (LECs). The assumption that the most important contributions to
the LECs come from the dynamics of the low-lying resonances, often referred to as the resonance saturation hypothesis,
has stimulated the use of large-$N_C$ resonance lagrangians in order to obtain explicit values for the LECs. We study the validity of the resonance saturation assumption at the next-to-leading order
in the $1/N_C$ expansion within the framework of Resonance
Chiral Theory (R$\chi$T). We find that,
by imposing QCD short-distance constraints, the chiral couplings can be written in terms of the resonance masses
and couplings and do not depend explicitly on the coefficients of the chiral operators in the Goldstone boson sector of
R$\chi$T. As we argue, this is the counterpart formulation of the resonance saturation statement in the context
of the resonance lagrangian. Going beyond leading order in the $1/N_C$ counting
allow us to keep full control of the renormalization scale dependence of the LEC estimates.
\end{abstract}

\pacs{11.15.Pg, 12.38.-t, 12.39.Fe}

\maketitle


\section{Introduction}

Chiral Perturbation Theory ($\chi$PT) is the effective field theory of
Quantum Chromodynamics (QCD) at very-low energies~\cite{ChPT}.
Theoretical calculations in this framework have nowadays reached the two-loop level accuracy.
The predictions from $\chi$PT are parameterized in terms of the $\cO(p^4)$ and $\cO(p^6)$ low-energy constants (LECs), which
are not fixed by the chiral symmetry.
Given the large uncertainties in the $\cO(p^4)$ coupling estimates~\cite{reviewChPT},
and the poor knowledge and huge number of $\cO(p^6)$ couplings
($90$ in $SU(3)$ and $53$ in $SU(2)$~\cite{p6ChPT}, just in the even intrinsic parity sector), 
a proper determination of LECs
turns into a difficult task.
Actually, one of the major problems for doing phenomenology
using $\chi$PT at the two-loop level
comes from our ignorance on these constants~\cite{reviewp6ChPT}.

Different theoretical approaches to determine the chiral couplings have
been pursued recently (see for instance  Ref.~\cite{Necco} for the state of
the art in the context of lattice QCD or the recent work~\cite{Prades} in the
framework of QCD sum-rules). Among the latter, the use of large-Nc
lagrangians has been proven very fruitful~\cite{RChTa,RChTb,RChTc,RchTGreen}.
By integrating out the resonances fields, one can obtain
the LECs parameterized in terms of resonance masses and couplings. The method
relies on the assumption that the most important contribution
to the LECs come from the physics of the low-lying resonances.

Resonance Chiral Theory (R$\chi$T)~\cite{RChTa,RChTb,RChTc} provides a chiral invariant lagrangian
description of the meson interactions in the resonance region,
$M_\rho \lsim E \lsim 2$~GeV, where the main problems to
develop a formal effective field theory are
the existence of many resonances with close masses and 
the apparent absence of a natural expansion parameter.
The phenomenological lagrangian is nevertheless ruled by the
$1/N_C$ expansion~\cite{largeNC}:
tree-level interactions between the large-$N_C$  infinite spectrum
of narrow states provide the leading order (LO)
contribution to Green functions of QCD currents,
being the next-to-leading order (NLO) corrections given by one-loop diagrams and subleading
tree contributions

The Green functions generated from the resonance lagrangian must obey QCD short-distance constraints.
The matching between Green functions evaluated from R$\chi$T~\cite{RchTGreen}
or constructed using general meromorphic functions with resonance poles~\cite{MHA,Pade,GF,saturation1}
and the asymptotic behaviour dictated by QCD yields valuable information on the
resonance couplings.
In this way Green functions describing the resonance region are used as a bridge between  QCD and ChPT,
allowing the determination of LECs in terms of a few hadronic parameters.
This matching procedure has been realized at the practical level by
approximating the hadronic spectrum to a finite number of states, thus
introducing a model-dependence in the description.
The truncation of the tower and the choice of an appropriate set of short-distance
constraints for each case constitute the so-called Minimal Hadronic Approximation~\cite{MHA},
which can be implemented in an equivalent way by using general meromorphic functions
or a chiral resonance lagrangian. An implicit hypothesis in the predictions obtained from R$\chi$T is
that they should approach the actual values of the QCD low-energy constants
as more and more resonances are added to the theory. This hypothesis is
non-trivial and has been a subject of investigation in
Refs.~\cite{saturation1,SanzCillero:2007ib,Golterman}. In particular, possible conflicts with the short-distance matching have
been pointed out in Ref.~\cite{Golterman}.

A systematic study of the large-$N_C$ determinations of the $\cO(p^4)$ and
$\cO(p^6)$ $\chi$PT couplings within R$\chi$T has been undertaken in
Refs.~\cite{RChTa} and~\cite{RChTc}, respectively.
A well-known drawback of the estimates obtained upon integration
of the resonances at tree-level is that
we are unable to control their renormalization scale dependence, being the latter
a subleading effect in $1/N_C$. Typically, these LO
predictions are assumed to correspond to a value $\mu\sim M_\rho$,
but a large uncertainty from variations of the scale is unavoidable.
Clearly, this arbitrariness on the choice of the scale for the
LEC estimates disappears if the matching between $\chi$PT and R$\chi$T is done
at the next-to-leading order in $1/N_C$ in both theories ({\it i.e.} 
including loops with mesons),
as it has been corroborated in recent attempts to determine the LECs
at NLO within R$\chi$T~\cite{cata,juanjo1,juanjo2,juanjo3}.

The estimate for the constant $L_i$ of a $\chi$PT operator ${\cal O}_i$
obtained from R$\chi$T depends on an equivalent (in principle unknown) constant $\widetilde{L}_i$,
corresponding to the coupling of an operator with the same structure as ${\cal O}_i$, but
living in the theory where the resonances are active degrees of freedom.
At LO, it was found in Ref.~\cite{RChTb} that the
couplings $\widetilde{L}_i$ corresponding to operators of chiral
order $\cO(p^4)$ were fixed in terms of resonance couplings and masses
once short-distance QCD constraints were
imposed in the effective lagrangian.
In particular, they were found to be zero in the antisymmetric tensor formalism
for spin--1 fields~\cite{RChTb}.
Therefore, upon tree-level integration of the resonances, one obtains
predictions for the $\chi$PT low-energy couplings
in terms of just resonance parameters related to operators which
involve resonance fields. The later statement provides a precise definition of
what should be understood by {\it resonance saturation} in the context of R$\chi$T.
It is the aim of this work to show that the statement also holds when
NLO corrections in $1/N_C$ in R$\chi$T are considered: 
the $\widetilde{L}_i$ get as well fixed in terms of
resonance parameters and, consequently, the prediction 
for the LECs only depend on the values of
the resonance couplings and masses.
This result has been used implicitly in recent works~\cite{juanjo2,juanjo3},
where NLO estimations of the constants $L_{8},\,C_{38}$ and $L_{10},\,C_{87}$ were extracted
from the analysis of the $SS-PP$ and the $VV-AA$ correlators, respectively.
The present paper is intimately related to Ref.~\cite{PRD}, which
was devoted to the same subject considering a R$\chi$T lagrangian
with just Goldstones, scalar and pseudoscalar resonances~\cite{oneloop}.
Here we provide the general proof including also vector and
axial-vector mesons, thus clarifying the role of resonance saturation in
a wider range of applications.

The outline of the paper is as follows.
In Section~\ref{sec:RchiT} we introduce the aspects of $\chi$PT and R$\chi$T which are relevant for our case,
focusing on the structure of the lagrangians and their related couplings and power-counting.
In Section~\ref{sec:saturation} we discuss the precise meaning of resonance saturation
in the framework of R$\chi$T, and give specific examples at leading
and next-to-leading order in $1/N_C$.  In Section~\ref{sec:NLOLECs}
we prove that resonance saturation is fulfilled for those
LECs related to QCD amplitudes obeying high-energy constraints.
The demonstration will be based on a careful
analysis of the analytic structure of the matrix elements
calculated with R$\chi$T up to NLO. Finally, Section~\ref{sec:summary}
summarizes our results.

\section{Chiral Resonance Lagrangian}
\label{sec:RchiT}

Chiral Perturbation Theory is organized as a perturbative expansion in powers of light quark masses and derivatives of the Goldstone fields~\cite{ChPT},
\begin{eqnarray}
\mathcal{L}_{\chi PT} &=& {\displaystyle \sum_{n \geqslant 1}} \mathcal{L}_{2n}^{\chi PT} \,, \label{chptlagrangian}
\end{eqnarray}
with $\mathcal{L}_{2n}^{\chi PT}\sim {\cal O}(p^{2n})$.
The leading-order term
\begin{eqnarray}
\mathcal{L}_2^{\chi PT} &=& \frac{F^2}{4} \bra u_\mu u^\mu + \chi_+ \ket \,
\end{eqnarray}
contains only two couplings, the meson decay constant in the chiral limit $F$ and the constant $B_0$
inside $\chi_+ \sim m_q B_0$, related to the quark condensate. The chiral tensor $u_\mu$ contains a derivative acting on the Goldstone fields so it is of order $p$ in the chiral counting. The tensor $\chi_+ \sim {\cal O}(M_{\pi}^2)$ counts as
order $p^2$ in the
standard formulation of $\chi$PT where the light quark mass and the momenta are related.
The low-energy constants in the effective lagrangian are not fixed by symmetry requirements and
their number increases quickly with the chiral order. At ${\cal O}(p^4)$ ten additional couplings are
allowed by the chiral symmetry:
\begin{align} \label{ChPTp4}
\mathcal{L}_4^{\chi\mathrm{PT}}& \!\!= L_1 \bra u_\mu u^\mu \ket^2 \!+\! L_2 \bra u_\mu u^\nu \ket \bra u^\mu u_\nu \ket \!+\! L_3 \bra u_\mu u^\mu u_\nu u^\nu \ket \nonumber \\ &+ L_4 \bra u_\mu u^\mu \ket \bra \chi_+ \ket
 +  L_5 \bra u_\mu u^\mu \chi_+ \ket + L_6 \bra \chi_+ \ket^2 \nonumber \\ &+ L_7 \bra \chi_-\ket^2
+ L_8/2 \, \bra \chi_+^2 + \chi_-^2 \ket
-i L_9 \bra f^{\mu\nu}_+ u_\mu u_\nu \ket \nonumber \\ & +L_{10}/4 \, \bra f_{+ \mu\nu}f_+^{\mu\nu}-f_{-\mu\nu}f_-^{\mu\nu}\ket  \,,
\end{align}
where the $SU(3)$ case has been considered and we have dismissed contact terms and operators that vanish when the equations of motion are used. We have showed explicitly the form of the ${\cal O}(p^4)$ chiral structures to identify the
relevant Green functions to which the LECs $L_i$ contribute. Since the vector, axial-vector, scalar and
pseudoscalar sources are contained in the chiral tensors $f_{+}^{\mu \nu},\,f_{-}^{\mu \nu},\,\chi_{+}$ and
$\chi_{-}$, respectively, and $u_\mu$ involves at least one Goldstone boson,
it follows that at $\cO(p^4)$ in the chiral limit:
(i) $L_1$, $L_2$ and $L_3$ determine the Goldstone boson scattering,
(ii) $L_4$ and $L_5$ the scalar form factor of the pion,
(iii) $L_6+L_7$ and $L_8$ the difference of the scalar and pseudoscalar correlators,
(iv) $L_6$ the two-point Green function of two scalar densities
$\bar{q} q$  and  $\bar{q}' q'$   with $q\neq q'$,
(v) $L_9$ the vector form factor of the pion,
and (vi) $L_{10}$ the difference of the two-point correlation function
of  vector and axial-vector currents.

For ${\cal O}(p^6)$ accuracy we have to account for 90 new independent terms only in the even intrinsic parity sector with coefficients $C_i$,
\begin{equation}
\mathcal{L}_6^{\chi\mathrm{PT(even)}} \,=\,  \sum_{i=1}^{90} C_i \cO_i^{(6)}\,.
\end{equation}

\hspace{0cm}From an effective field theory point of view, the Goldstone
interactions at low-energies are affected by the dynamics of hadronic
states of higher masses (resonances), which have been integrated out.
These effects can be studied
systematically with the help of a lagrangian
description of the chiral invariant Goldstone-resonance interactions which
takes the $1/N_C$ expansion as a guiding principle, as described in Refs.~\cite{RChTa,RChTb,RChTc}.
The lagrangian of Resonance Chiral Theory can be organized according to the number of resonance fields
in the interaction terms,
\begin{eqnarray}
\mathcal{L}_{\mathrm{R}\chi\mathrm{T}}&=& \mathcal{L}^{\mathrm{GB}} + \mathcal{L}^{R_i} + \mathcal{L}^{R_iR_j} + \mathcal{L}^{R_iR_jR_k}
+\dots \,, \label{rchtlagrangian}
\end{eqnarray}
where $R_i$ stands for resonance multiplets of vectors $V(1^{--})$,
axial-vectors $A(1^{++})$, scalars $S(0^{++})$ and
pseudoscalars $P(0^{-+})$.
In order to carry out the matching of R$\chi$T and $\chi$PT,
one is forced to truncate the infinite tower of resonances
of the large-$N_C$ spectrum to a finite amount of them.
As more hadrons are added to the theory,
the  R$\chi$T amplitudes are expected to approach progressively 
its actual large--$N_C$ value~\cite{MHA,Pade}.
The arguments that will be used to establish resonance saturation in the next
sections can be easily generalised to any number of multiplets. To simplify the
discussion we will nevertheless assume in the following a large-$N_C$ lagrangian
${\cal L}_{\mathrm{R}\chi\mathrm{T}}$ truncated to the lowest-lying resonance
multiplet in each channel.

The term ${\cal L}^{\mathrm{GB}}={\cal L}^{\mathrm{GB}}_2+{\cal L}^{\mathrm{GB}}_4+\dots$ in
Eq.~(\ref{rchtlagrangian}) is the Goldstone chiral lagrangian,
which shares the same general
operator structure as the $\chi$PT lagrangian~(\ref{chptlagrangian}).
However, the values of the ${\cal L}^{\mathrm{GB}}$ couplings  
($\widetilde{L}_i,\,\widetilde{C}_i\dots$) are different from the 
$\chi$PT LECs ($L_i,\,C_i\dots$),
since the Goldstone self-interactions in the presence of resonances differ from 
those at low-energies where the resonances have been integrated out. 
Therefore a new set of (a priori unknown) constants is introduced in
the Goldstone sector of the resonance lagrangian.

Adopting the antisymmetric tensor formalism to describe the spin-1 resonance fields,
the second term in Eq.~(\ref{rchtlagrangian}) reads~\cite{RChTa}
\begin{align}
\mathcal{L}^{R_i} &=\, \mathcal{L}^V + \mathcal{L}^A + \mathcal{L}^S + \mathcal{L}^P \, ,  \phantom{\frac{1}{2}} \nonumber \\
\mathcal{L}^V_{(2)} &=\, \frac{F_V}{2\sqrt{2}} \bra V_{\mu\nu} f^{\mu\nu}_+ \ket \,+\, \frac{i\, G_V}{2\sqrt{2}} \bra V_{\mu\nu} [u^\mu, u^\nu] \ket \, ,  \nonumber \\
\mathcal{L}^A_{(2)} &=\, \frac{F_A}{2\sqrt{2}} \bra A_{\mu\nu} f^{\mu\nu}_- \ket\, , \nonumber\\
\mathcal{L}^S_{(2)} &=\, c_d \bra S u_\mu u^\mu\ket\,+\,c_m\bra S\chi_+\ket\,  \, , \phantom{\frac{1}{2}}   \nonumber \\
\mathcal{L}^P_{(2)} &=\, i\,d_m \bra P \chi_- \ket \phantom{\frac{1}{2}}  , \label{1Rlagrangian}
\end{align}
where only the terms with a chiral tensor of ${\cal O}(p^2)$ have been shown.
The lagrangian
(\ref{1Rlagrangian}), supplemented with the kinetic and mass terms
for the resonance fields, yields the most general
lagrangian that can give contributions to the ${\cal O}(p^4)$ LECs
after integrating out the resonances at tree-level (\textit{i.e.}
at LO in the large-$N_C$ expansion)~\cite{RChTa,RChTb}.
The latter is easily understood because integrating out a resonance field $R_i$
yields a series of chiral monomials starting at
${\cal O}(p^2)$. Thus for the
chiral ${\cal O}(p^6)$ LECs, the terms in ${\cal L}^{R_i}$, ${\cal L}^{R_iR_j}$ and
${\cal L}^{R_iR_jR_k}$ with chiral tensors of order $p^4,\,p^2$ and $p^0$,
respectively, are required (the complete basis of operators can be found in Ref.~\cite{RChTc}). This counting
does not necessarily apply at NLO in $1/N_C$ because resonances occur inside loops.
The loop integration can trade off powers of $p^2$ for resonance masses, so that
the low-energy expansion of such contributions can yield monomials in $p^2$ of
lower order than expected by dimensional analysis. Thus
resonance interactions with
higher derivatives that do not contribute to ${\cal O}(p^4)$ and ${\cal O}(p^6)$ LECs upon integration
at tree-level can become relevant when loop effects are taken into account. Since our analysis
concerns the determination of the LECs including one-loop contributions in the resonance chiral theory, we
shall formally consider resonance terms in ${\cal L}_{\mathrm{R}\chi\mathrm{T}}$ with chiral tensors of arbitrary order.
Nevertheless, for most phenomenological applications
the interactions with large number of derivatives in Eq.~(\ref{rchtlagrangian}) can be shown
to violate the QCD ruled asymptotic behavior of Green functions and form factors 
at high energies, and thus get severely restricted.

\section{Resonance saturation in R$\chi$T}
\label{sec:saturation}

The notion of resonance saturation has been vaguely used
in the literature of low-energy QCD to
designate a number of cases where the strong interactions
are essentially described by meson-resonance exchanges.
In the framework of the large-$N_C$-inspired lagrangian
of Eq.~(\ref{rchtlagrangian}),
resonance saturation is linked to the estimation of
the chiral LECs from the knowledge
of the resonance parametes.  In particular, it is related
to the role of the couplings in the Goldstone sector
$\mathcal{L}^{\mathrm{GB}}$, as we argue next.

Upon integration of the resonances one gets an expression for any chiral coupling
in terms of the parameters in the R$\chi$T lagrangian:
\begin{eqnarray}
L_i&=&\widetilde{L}_i + f_i(M_{R},\alpha_R) \,,\nonumber \\
C_i&=&\widetilde{C}_i + g_i(M_{R},\alpha_R) \,, \dots   \label{saturation}
\end{eqnarray}
where $f_i(M_{R},\alpha_{R})$ and $g_i(M_{R},\alpha_{R})$ are the contribution stemming from the low-energy expansion of the resonance contributions
({\it i.e.} from the diagrams that contain resonance lines), which include one-loop diagrams if we work at NLO in $1/N_C$. $M_R$ denotes generically the resonance masses while
$\alpha_R$ stands for the R$\chi$T couplings accompanying operators with resonance fields. Clearly, Eq.~(\ref{saturation})
is useless for determining the LECs if the couplings $\widetilde{L}_i$ in the Goldstone boson sector of ${\cal L}_{\mathrm{R}\chi\mathrm{T}}$ remain unknown parameters. We shall state that resonance saturation is fulfilled
in the matching between ${\cal L}_{\mathrm{R}\chi\mathrm{T}}$ and ${\cal L}_{\chi\mathrm{PT}}$
if the $\widetilde{L}_i$ couplings get fixed completely by the short-distance constraints,
so that the $L_i$ are then given as functions of only $M_R$ and $\alpha_R$. This definition implies that
the saturation is accomplished for any value of the $\chi$PT renormalization scale $\mu$ (the ``extreme'' version of resonance saturation pointed out in Ref.~\cite{cata}).

The condition of resonance saturation was confirmed at leading-order in $1/N_C$ for the ${\cal O}(p^4)$ LECs in Refs.~\cite{RChTa,RChTb}. In the R$\chi$T formulation where spin-1 resonances are described by antisymmetric
tensor fields it was found that the $\widetilde{L}_i$ actually vanish due to short-distance constraints and Eq.~(\ref{saturation}) turns out to be
\begin{align}
&L_1=\frac{G_V^2}{8M_V^2}\,,\quad
L_2=\frac{G_V^2}{4M_V^2}\,, \quad
L_3=-\frac{3G_V^2}{4M_V^2}\,+\,\frac{c_d^2}{2M_S^2}\, ,\nonumber \\
&L_5=\frac{c_dc_m}{M_S^2}\,,\quad
L_8=\frac{c_m^2}{2M_S^2}\,-\,\frac{d_m^2}{2M_P^2} \,, \quad
L_9=\frac{F_VG_V}{2M_V^2}\,, \nonumber \\
&L_{10}=-\frac{F_V^2}{4M_V^2}\,+\,\frac{F_A^2}{4M_A^2}\,,\qquad
L_4=L_6=L_7=0\,, \label{rescont}
\end{align}
that is, one is able to determine the $\cO(p^4)$ chiral couplings of Eq.~(\ref{ChPTp4}) in terms of the resonance parameters in Eq.~(\ref{1Rlagrangian}). For the $\cO(p^6)$ LECs $C_i$ a complete analysis of the saturation hypothesis has not been yet performed but it has often been assumed to hold in phenomenological applications. (For a recent review of the status of the LEC determinations from R$\chi$T see Ref.~\cite{Pich:2008xj} and references therein).

As an illustrative example of how the saturation is fulfilled at leading order
in $1/N_C$, let us consider the two-point correlation functions of two vector
or axial-vector currents in the chiral limit. Of particular interest is their
difference $\Pi(q^2)\equiv \Pi_{VV}(q^2)-\Pi_{AA}(q^2)$.
At large $N_C$, the resonance lagrangian of Eq.~(\ref{rchtlagrangian}) produces for the
$VV-AA$ charged correlator~\cite{juanjo3}
\begin{equation}
\Pi(q^2)=\frac{2F^2}{q^2} + \frac{2F_V^2}{M_V^2-q^2} - \frac{2F_A^2}{M_A^2-q^2}
- 8\widetilde{L}_{10}  +
16  \widetilde{C}_{87} \,q^2
+... \label{VA_LO_RChT}
\end{equation}
being $q$ the momentum flowing into the current vertex. In Eq.~(\ref{VA_LO_RChT})
we have explicitly shown  contributions from
$\mathcal{L}^{\mathrm{GB}}$ up to chiral order $p^6$. From QCD we know that
this correlator vanishes when $q^2\rightarrow \infty$~\cite{VA}. Imposing this behavior in Eq.~(\ref{VA_LO_RChT}), one gets that $\widetilde{L}_{10}=\widetilde{C}_{87}=0$.
The same correlator computed at low-energies with the $\chi$PT lagrangian  gives
\begin{eqnarray}
\Pi^{\chi P T}(q^2)&=&\frac{2F^2}{q^2}  - 8 L_{10}  + 16  C_{87} \,q^2 +\, ...
 \label{VA_LO_ChPT}
\end{eqnarray}
Comparing this result with the low-energy expansion of Eq.~(\ref{VA_LO_RChT}),
one is able to estimate the corresponding chiral couplings as a function
of the resonance masses and couplings only:
\begin{eqnarray}
L_{10} =  -\frac{F_V^2}{4M_V^2}\,+\,\frac{F_A^2}{4M_A^2} \,,
\qquad
C_{87}= \frac{F_V^2}{8M_V^4}\,-\,\frac{F_A^2}{8M_A^4} \,.
\label{ChPTmatchingLO}
\end{eqnarray}
Note that the important point for the resonance saturation condition to hold is that the value of the corresponding
${\cal L}^{\mathrm{GB}}$ couplings have been fixed by requiring the right short-distance behaviour
of the Green function computed with the resonance theory. A fixed but non-zero value
for the ${\cal L}^{\mathrm{GB}}$ couplings can be required for consistency with QCD if a different representation for the
spin-1 resonance fields (the Proca formalism, for example) is chosen~\cite{RChTb}.
A similar situation has also been found in a recent analysis of spin--2 resonances~\cite{LECs-Ecker},
which shows that some of the coefficients in $\mL^{\rm GB}$ must be
different from zero, though
the $\chi$PT LECs get finally determined in terms of resonance parameters only.

The resonance saturation condition applies also at next-to-leading order
in the $1/N_C$ expansion. We consider again the case of the
$VV-AA$ correlator $\Pi(q^2)$ as an example. Its expression up to  NLO within R$\chi$T
can be split in the following way:
\begin{eqnarray}
\Pi(q^2) &=& \Pi^{\good}  (q^2 )+ \Pi^{\bad} (q^2 ) +\Pi^{\mathrm{GB}} (q^2)
\, .
\label{Pibaddef}
\end{eqnarray}
The term $\Pi^{\good} (q^2)$ collects the part of the amplitude which
vanishes at high energies, while
$\Pi^{\bad}(q^2)$ and $\Pi^{\mathrm{GB}}(q^2)$ collect the pieces which
grow as $\cO(q^0)$ or faster for large $q^2$. The piece
$\Pi^{\mathrm{GB}}(q^2)$ arises from the local contributions of $\mL^{\rm GB}$,
\begin{eqnarray}
\Pi^{\rm GB}(q^2)&=& - 8\, \widetilde{L}_{10}
\, +\,  16 \,  \widetilde{C}_{87}\,   q^2 + \dots
\end{eqnarray}
and only one-loop diagrams contribute to the piece $\Pi^{\bad} (q^2 )$, since
the tree-level meson exchanges already satisfy the short-distance constraints.
Resonance saturation states that the local terms
$\widetilde{L}_i ,\,\widetilde{C}_i \dots$
get fixed once the correct asymptotic behavior
is imposed in the theory. In the case at hand, the short-distance constraint implies that
\begin{eqnarray}
\Pi^{\bad} (q^2 ) +\Pi^{\mathrm{GB}} (q^2) &=& 0
\,,\label{solve}
\end{eqnarray}
and consequently
\begin{eqnarray}
\Pi  (q^2) &=& \Pi^{\good} (q^2) \, . \label{piNLO}
\end{eqnarray}

\hspace*{0cm}From the low-energy expansion of Eq.~(\ref{piNLO})
one can extract estimates for the renormalized chiral couplings
$L_{10}^r$, $C_{87}^r$ with NLO accuracy, thus keeping
control of their one-loop renormalization scale dependence: 
the matching equations between R$\chi$T and $\chi$PT read
\begin{eqnarray}
-8 \, L_{10}^r(\mu)&=& \lim_{q^2\to 0} \Big[\Pi^{\good} (q^2)
- \hat{\Pi}^{\chi\mathrm{PT}} (q^2;\mu) \Big]
\,, \nonumber \\
16\, C_{87}^r(\mu)&=& \lim_{q^2\to 0} \bigg[ \frac{\mathrm{d}}{\mathrm{d} q^2}\Big( \Pi^{\good} (q^2)
- \hat{\Pi}^{\chi\mathrm{PT}} (q^2;\mu) \Big)\bigg] ,
\nonumber\\
\label{ChPTmatchingNLO}
\end{eqnarray}
where $\hat{\Pi}^{\chi\mathrm{PT}}$ is the $VV-AA$ correlator
in $\chi$PT~\cite{ChPT} without the local contributions
$L_{10}^r$, $C_{87}^r$...,
which have been isolated on the l.h.s. of the equations above.
The cancellation of the $\ln{(-q^2)}$ terms in
the differences in the r.h.s of Eqs.~(\ref{ChPTmatchingNLO}) is ensured
because R$\chi$T reproduces $\chi$PT at low energies.
As a consequence, the limits are well defined for $q^2\to 0$  and
the $\mu$ dependence of $\hat{\Pi}^{\chi \rm PT}(q^2;\mu)$
gives the right renormalization scale running of the $\chi$PT LECs.

The solution of Eq.~(\ref{solve}) for the ${\cal L}^{\mathrm{GB}}$
couplings requires that $\Pi^{\mathrm{bad}} (q^2)$ does not
have non-polynomial terms  (like
$\log (-q^2)$). In Section~\ref{sec:2point},
we show  that for a general two-point correlator of two currents
the non-polynomial terms can be made to vanish by a suitable set of the tree-level
resonance parameters.
To extend the saturation condition to the
LECs related to the pion form factors and to Goldstone boson scattering one has to impose
in an analogous way that the parts with the wrong asymptotic
behaviour in the corresponding amplitudes can be made free of non-polynomial terms.
Sections~\ref{sec:3point} and ~\ref{sec:4point} deal with the proof for the pion form factors and
for Goldstone scattering, respectively.

Resonance saturation at NLO with a resonance lagrangian involving only scalars and pseudoscalars mesons was discussed in Ref.~\cite{PRD}. There it was found that for those LECs
which get tree-level contributions from scalar and
pseudoscalar resonance exchange (namely $L_{4-8}$),
the corresponding local couplings in R$\chi$T get fixed to $\widetilde{L}_i=0$. This more
restrictive condition arises because the analysis in Ref.~\cite{PRD} proved that the
short-distance conditions on the $SS$ and $PP$ correlators yield $\Pi^{\mathrm{bad}}(q^2)=0$, so
that $\widetilde{L}_i=\widetilde{C}_i=0$ follow from Eq.~(\ref{solve}).
A key ingredient in the proof of Ref.~\cite{PRD} was that the spin-0
resonance propagator
behaves as ${\cal O}(1/q^2)$ for large $q^2$. In this work
we address the more general case
which accounts also for spin-1 resonance fields. They are conventionally
described in R$\chi$T in the antisymmetric tensor field formalism though
other formalisms can be used~\cite{RChTa,RChTb}.
Contrary to the spin-0 case, the propagator of a
massive spin--1 particle scales like $\cO(q^0)$ for $q^2\to \infty$. Recall,
for instance, the form of the propagator in the Proca formalism,
\begin{eqnarray}
\Delta^{\mu\nu}(q^2) =
\frac{-i}{q^2-M_R^2}\,  \left(g^{\mu\nu} \, -\, \Frac{q^\mu q^\nu}{M_R^2}\right)
\,. \label{vector}
\end{eqnarray}
Similarly, the antisymmetric field propagator contains a piece that
does not fall off as $\cO(1/q^2)$ for $q^2\to \infty$ (see Ref.~\cite{RChTa} for the explicit expression).
Therefore, the
conclusions obtained for scalar and pseudoscalar resonances by
inspection of the large $q^2$ behaviour of the one-loop
amplitudes in Ref.~\cite{PRD} do not translate trivially
to the case of vectors and axial-vector resonances, which
require an independent analysis.

\section{QCD amplitudes at NLO in $1/N_C$}
\label{sec:NLOLECs}

In this section we analyze the asymptotic behaviour for
large momenta of the amplitudes which are relevant to fix the ${\cal L}^{\mathrm{GB}}$ couplings
through conditions analogous to Eq.~(\ref{solve}). In particular, we prove that non-polynomial terms
do not appear in the latter conditions when the amplitudes obey the short-distance constraints.

\begin{figure}[t]
\includegraphics[width=6cm]{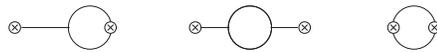}
\caption{Topologies in the one-loop correlators with an intermediate two-meson cut.
The lines can represent both Goldstone and resonance fields. \label{fig:correlators}}
\end{figure}

The leading-order term in the large-$q^2$ expansion of a R$\chi$T one-loop amplitude is obtained easily by simple dimensional
analysis. The large-$q^2$ counting rules are that any vertex with a chiral tensor of order $p^{(2n)}$ yields at most  $q^{2n}$. Spin-0 and spin-1 propagators outside the loop~\footnote{When the vector (axial-vector) resonance is connected to an external vector (axial-vector) current, the part of the spin--1 propagator which behaves as $\cO(q^0)$ for large $q^2$ does not contribute because of the structure of the vertices. The same holds when the
vector resonance is coupled to a pair of Goldstone bosons.} count as $(q^2)^{-1}$, while a spin-1 propagator inside a loop is ${\cal O}(q^0)$
(see Eq.~(\ref{vector})). The loop integration measure gives $\int d^4k \sim q^4$. The chiral tensors
containing the external currents ({\it i.e.} $\chi_{\pm}$ and $f_{\pm}^{\mu\nu}$) can be booked as ${\cal O}(q^0)$
once we drop the momentum tensor structure introduced by the vector and axial-vector currents.
As an example, the tree-level exchange of a vector resonance in the correlator of two vector currents ($\Pi_{VV}$) counts as $1/q^2$ according to the rules, which is clear from the explicit result:
\begin{equation}
\Pi_{VV}(q^2)=\frac{2F_V^2}{M_V^2-q^2}\,.
\end{equation}
One-loop corrections to $\Pi_{VV}$ arise from diagrams as those shown in Fig.~{\ref{fig:correlators}}. Let us assume that all vertices in the diagrams contain a chiral tensor of ${\cal O}(p^2)$.
If there are no spin-1 resonances running inside the loops then the
diagrams behave as ${\cal O}(q^0)$ for large $q^2$. If we allow for one or two spin-1 resonances in the loops then the diagrams are up to ${\cal O}(q^2)$ or ${\cal O}(q^4)$, respectively.

We analyse in what follows the large-$q^2$ structure of the two-current
correlators, pion form factors and Goldstone scattering amplitude separately.
To simplify the expressions we will consider only resonance operators in
${\cal L}^{R_i},\,{\cal L}^{R_i R_j}\,\dots$ with a chiral tensor up to ${\cal O}(p^2)$.
The generalization
of our findings for the case of higher-order interaction terms would be straightforward.

\subsection{Two-point correlators}
\label{sec:2point}

Let us consider the two-point functions built from two scalar ($SS$) or pseudoscalar
($PP$) densities, or vector ($VV$) or axial-vector ($AA$) currents.
Their tree-level expressions are given by one-particle exchanges, so they are booked as $\cO(q^{-2})$ at large energies according to the counting explained above.
On the other hand, the one-loop diagrams  can give up to $\cO(q^4)$ terms if we allow for spin-1 mesons in the absorptive part.
After reduction to scalar integrals, all one-loop terms are proportional
to the scalar two- and one-point functions $B_0(q^2,M_{i}^2,M_{j}^2)$
and $A_0(M_{i}^2)$~\cite{Passarino:1978jh}. Expanding out the coefficients
in front of the scalar integrals for $q^2\to \infty$, the
one-loop amplitude has the form~\cite{PRD,SanzCillero:2007ib}
\begin{eqnarray}
&&\Pi^{\mathrm{1-loop}}(q^2)=
\nonumber\\
&&\hspace*{0.75cm}\sum_{n} B_0(q^2,M^2_i,M^2_j)\left( \lambda^{(0)}_{n} q^0  + \lambda^{(2)}_{n} q^2 +  \lambda^{(4)}_{n} q^4  \right)
\nonumber \\
&& \hspace*{0.75cm}+ \sum_\ell  A_0(M_\ell^2) \left( \beta^{(0)}_\ell  q^0 + \beta^{(2)}_\ell q^2 + \beta^{(4)}_\ell q^4  \right)
\nonumber \\
&& \hspace*{0.75cm}+ \left( \gamma^{(0)} q^0 + \gamma^{(2)}q^2 + \gamma^{(4)} q^4 \right)
\,\,\,+\cO\left(\frac{1}{q^2}\right)  \,,
\label{correlators}
\end{eqnarray}
where the sum in $n$ extends to all pairs of virtual mesons with masses $M_i$, $M_j$ that occur in the loops, and $\lambda^{(2k)}_{n}$, $\beta^{(2k)}_{\ell}$ and $\gamma^{(2k)}$ are combinations of resonance parameters. The superindex
$(2k)$ refers to the order at large-$q^2$ of the corresponding term.
If we further expand the scalar functions around $q^2=\infty$, the one-loop amplitude shows
the general analytical structure
\begin{eqnarray}
\Pi^{\mathrm{1-loop}}(q^2) &=& \left( \hat{\lambda}^{(0)} q^0
+ \hat{\lambda}^{(2)}q^2+\hat{\lambda}^{(4)}q^4    \right) \,\,\,
\ln{\Frac{-q^2}{M_R^2}}\,
\nonumber \\
&& \hspace*{-1.25cm}
+ \left( \hat{\gamma}^{(0)}  q^0 + \hat{\gamma}^{(2)} q^2  + \hat{\gamma}^{(4)} q^4   \right)
\,\,\,+\cO\left(\frac{1}{q^2}\right)  ,
\label{correlators2}
\end{eqnarray}
with $M_R$ some arbitrary mass scale chosen to make the argument of the logarithms
dimensionless. Note that the logarithmic part contains the absorptive contributions which
define the spectral function $\mathrm{Im}\, \Pi(q^2)$.

Local terms from $\mathcal{L}^{\mathrm{GB}}$ also contribute to the correlators
through a polynomial in the $\widetilde{L}_i,\,\widetilde{C}_i\dots$ couplings:.
\begin{eqnarray}
\Pi^{\rm{GB}}(q^2) &=&  \widetilde{L}_J\,
+   \widetilde{C}_J \,q^2 +
\dots
\label{GBpolynomial}
\end{eqnarray}
where the $\widetilde{L}_J$, $\widetilde{C}_J$... refer
to corresponding LECs or combination of them for the amplitude under consideration.
As mentioned in Section~\ref{sec:RchiT}, the relevant
${\cal O}(p^4)$ couplings  for the  correlators in the chiral limit are
$\widetilde{L}_{6-8}$ and $\widetilde{L}_{10}$. The ${\cal L}^{\mathrm{GB}}$
couplings in Eq.~(\ref{GBpolynomial}) should be understood as bare parameters,
which absorb the local ultraviolet divergences
that may be contained in the $\hat{\gamma}^{(2k)}$.

The amplitudes for the linear combinations of correlators $SS-PP$ and $VV-AA$ also have the
form shown in Eqs.~(\ref{correlators},\,\ref{correlators2}).
These correlators are particularly useful for the
purposes of determining the LECs, since we know they must vanish for $q^2\to \infty$,
as dictated by perturbative QCD~\cite{VA}.
This requirement translates into conditions on the terms shown
in Eq.~(\ref{correlators2}), which have the wrong short-distance
behaviour. Due to their different analytical structure, the cancellations must occur
separately for the logarithmic and polynomial parts.
The vanishing of the non-polynomial  part
requires that $\hat{\lambda}^{(2k)}=0$.
The cancellation of the remaining polynomial is then achieved
by tuning the local contributions from $\mL^{\rm GB}$ to fulfill the equations
\begin{eqnarray}
\widetilde{L}_J\,  +   \hat{\gamma}^{(0)}  =0 \, ,
\qquad
 \widetilde{C}_J\,  +   \hat{\gamma}^{(2)}  = 0 \, ,\quad \dots
 \label{LGBfixing}
\end{eqnarray}
These constraints  fix the value of the
corresponding ${\cal L}^{\mathrm{GB}}$ couplings that contribute to
the $VV-AA$ and $SS-PP$ correlators.  Thus, 
the piece with the wrong high-energy behaviour
disappears from the calculation and
becomes  irrelevant for the matching with $\chi$PT at low-energies.

Let us mention that a more restrictive condition than $\hat{\lambda}^{(2k)}=0$
has been used in the literature~\cite{juanjo2,juanjo3} to enforce the right short-distance behaviour.
The spectral function $\mathrm{Im}\, \Pi(q^2)$ associated to the $SS$, $PP$, $VV$ and $AA$ correlators are the sum of absorptive contributions corresponding to the different intermediate states,
\begin{equation}
\mathrm{Im}\, \Pi(q^2) = \sum_n \mathrm{Im}\, \Pi_n(q^2) \,.
\end{equation}
At one-loop, any of the possible absorptive contributions, $n$, comes from the two-particle cuts in the diagrams of Fig.~\ref{fig:correlators}.
At large $q^2$ the vector and axial-vector spectral functions tend to a constant whereas the scalar and pseudoscalar ones grow like $q^2$~\cite{VA}. Therefore, since there is an infinite number of possible states, the absorptive contribution in the spin-$1$ correlators coming from each intermediate state should vanish in the $q^2\rightarrow \infty$ limit if we assume a similar short-distance behavior for all of them. The high-energy behavior of the spin-$0$ spectral functions
is not so clear as, {\it a priori}, a constant behavior for each intermediate cut does not seem to be excluded.
However, the fact that $\Pi_{SS}(q^2)-\Pi_{PP}(q^2)$ vanishes as $q^{-4}$~\cite{VA}, the Brodsky-Lepage rules for the form factors~\cite{BrodskyLepage} and the $q^{-2}$ behavior of each one-particle intermediate cut seems to point out that every absorptive contribution to $\mathrm{Im}\, \Pi(q^2)$ must also vanish at large momentum transfer. This assumption translates into $\lambda^{(2k)}_n=0$ in Eq.~(\ref{correlators}), which is a particular solution of the more general condition
$\hat{\lambda}^{(2k)}=0$ pointed out above.

\subsection{Pion form factors}
\label{sec:3point}

\begin{figure}[t]
\includegraphics[width=5.5cm]{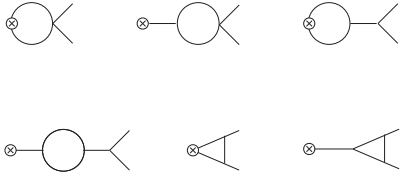}
\caption{Topologies in the one-loop pion form factor with a two-meson absorptive cut.
\label{fig:formfac}}
\end{figure}

We now turn to the analysis of the saturation condition for the LECs related to the scalar form
factor ($L_4$ and $L_5$ at ${\cal O}(p^4)$), and with the vector form factor of the pion ($L_9$).
The tree-level expression of the pion form factors behaves now as $\cO(q^0)$ at large energies,
while one-loop corrections given by the R$\chi$T lagrangian are up to $\cO(q^8)$.
The allowed topologies with absorptive two-meson cuts are shown in Fig.~\ref{fig:formfac}. The main difference with respect
the analysis for the two-point correlators comes from the presence of triangle graphs.
After the reduction of the one-loop diagrams to scalar functions, the form factor in the $q^2\to\infty$
expansion reads
\begin{align}
&\mathcal{F}^{\mathrm{1-loop}}(q^2) =
\nonumber\\ &
\quad \sum_n \sum_{\ell} C_0(q^2,0,0,M_i^2,M_j^2,M_\ell^2)\,
\left( \kappa_{n,\ell}^{(2)} \,q^4 
\right)
\nonumber \\ &
\quad + \! \sum_{n}
B_0(q^2,M^2_i,M^2_j)\!\left( \lambda_n^{(2)} q^2  \!+\!
\lambda_n^{(4)}q^4 \!+\!  \lambda_n^{(6)} q^6 \!+\!  \lambda_n^{(8)} q^8  \right)
\nonumber\\ &
\quad +\sum_\ell  A_0(M_\ell^2) \left( \beta^{(2)}_\ell  q^2 \! +\!
\beta^{(4)}_\ell q^4 \!+\!\beta^{(6)}_\ell  q^6 \! +\!
\beta^{(8)}_\ell q^8 \right)
\nonumber \\ &
\quad+ \left(  \gamma^{(2)} q^2 \!+\! \gamma^{(4)}q^4
\!+\!  \gamma^{(6)} q^6 \!+\!  \gamma^{(8)} q^8   \right)
+\cO\left(q^0 \right) . \label{formfactors}
\end{align}
The sum over $n$ extends to every $s$-channel two-particle cut
in the diagrams of Fig.~\ref{fig:formfac} involving
mesons $M_i$ and $M_j$. The three-point functions $C_0$ are generated from the
triangle diagrams with internal masses
$M_{i,j,\ell}$, where the subindex `$\ell$' labels the virtual meson connecting
the two outgoing pions.
As before, the quantities
$\kappa_{n,\ell}^{(2)}$, $\lambda^{(2k)}_n$, $\beta^{(2k)}_\ell$ and $\gamma^{(2k)}$
are combinations of resonance parameters (which obviously differ from those in Eq.~(\ref{correlators}),
although we are using the same notation). It is straightforward to check that
terms  of order $q^4$, $q^6$ and $q^8$ proportional to  $C_0$ do not show up if we stick to resonance
interaction terms with a chiral tensor of at most $\cO(p^2)$. Note that since $C_0$ behaves
asymptotically as $1/q^2$, the term $\kappa_{n,\ell}^{(2)}$ is of $\cO(q^2)$ for large $q^2$.

The expansion of the scalar functions at high energies leads to the analytic structure
\begin{align}
\mF&^{\mathrm{1-loop}}(q^2) =
\,\,\left( \hat{\kappa}^{(2)}q^2\right)
\,\,\,
\ln^2{\Frac{-q^2}{M_R^2}}\,
\nonumber \\
& \quad
+ \left( \hat{\lambda}^{(2)} q^2
+ \hat{\lambda}^{(4)}q^4+\hat{\lambda}^{(6)}q^6 +\hat{\lambda}^{(8)}q^8    \right) \,\,\,
\ln{\Frac{-q^2}{M_R^2}}\,
\nonumber \\
& \quad
+ \left( \hat{\gamma}^{(2)}  q^2 + \hat{\gamma}^{(4)} q^4  + \hat{\gamma}^{(6)} q^6 + \hat{\gamma}^{(8)} q^8   \right)
+\cO\left(q^0\right)  ,
\label{FF2}
\end{align}
with $M_R$ some arbitrary mass scale chosen to make the argument of the logarithms
dimensionless. The log squared terms arise from the $C_0$ functions.

The contributions to the form factor from ${\cal L}^{\rm{GB}}$ start now at ${\cal O}(q^2)$:
\begin{eqnarray}
\mathcal{F}^{\mathrm{GB}}(q^2) &=&
\Frac{\widetilde{L}_J\, q^2}{F^2}
+ \Frac{\widetilde{C}_J \, q^4}{F^2} +
\dots  \,,
\nonumber\\
\label{GBpolynomialF}
\end{eqnarray}
where, again, $\widetilde{L}_J$ and $\widetilde{C}_J$ refer to the
corresponding combination of the $\cO(p^4)$ and $\cO(p^6)$ LECs, respectively.
As an example, for the vector form factor one has 
$\widetilde{L}_J=2 \widetilde{L}_9$ and $\widetilde{C}_J= 4 \widetilde{C}_{90} - 4 \widetilde{C}_{88}$. 

Based on the Brodsky-Lepage rules for the form factors of QCD currents~\cite{BrodskyLepage},
and on the large-$q^2$ behaviour of the spin-0 and spin-1 current correlators (${\cal O}(q^2)$ for the spin-0
and ${\cal O}(q^0)$ for the spin-1 case) obtained from perturbative QCD,
we can expect that the
pion form factors behave at worst as a constant for $q^2\to \infty$.
This short-distance constraint requires that the terms shown in
Eqs.~(\ref{FF2},\,\ref{GBpolynomialF})
vanish when put together.
The vanishing of the logarithmic terms implies the
constraints $\hat{\lambda}^{(2k)}=\hat{\kappa}^{(2k)}=0$, and already lead
to a well behaved spectral function $\mathrm{Im}\, \mF(q^2)$.

As a result, only polynomial terms remain in Eq.~(\ref{FF2}),
which must be canceled with the local terms in Eq.~(\ref{GBpolynomialF})
by a suitable choice of the couplings of $\mathcal{L}^{\mathrm{GB}}$.
The procedure leads to equations analogous to those shown in Eq.~(\ref{LGBfixing}).
In this way, the expressions for the  R$\chi$T vector and scalar pion form factors with
the proper short-distance behaviour do not longer depend
on the  $\widetilde{L}_i,\,\widetilde{C}_i\dots$ couplings.
Thus, the related $\chi$PT couplings
($L_4$, $L_5$ and $L_9$ at $\cO(p^4)$) get determined in terms of the resonance parameters
at NLO in $1/N_C$ through  a matching with the R$\chi$T result at low-energies.

As in the case of the two-current correlators, it is possible
to consider the more stringent constraints which follow from
requiring that each individual contribution to the spectral function $\mathrm{Im}\, \mF(q^2)$ vanish
at high energies. This is achieved by the conditions
$\lambda^{(2k)}_n=\sum_\ell\kappa^{(2k)}_{n,\ell}=0$ for any $s$-channel absorptive cut $n$.

We want to emphasize that Sections~\ref{sec:2point} and ~\ref{sec:3point}
generalize the results of Ref.~\cite{PRD} to the case when  both spin-0 and spin-1
resonance fields are present in R$\chi$T. 

\subsection{Goldstone boson scattering}
\label{sec:4point}

\begin{figure}[t]
\includegraphics[width=6.5cm]{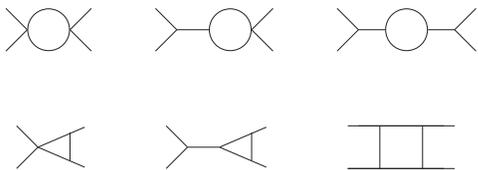}
\caption{Topologies in the one-loop scattering amplitude
with two-meson cuts in the $s$, $t$
or $u$--channels.
The possible permutations in the labeling of the external legs are implicitly
assumed.
\label{fig:scat}}
\end{figure}

The $2\to 2$ Goldstone boson scattering amplitude allows us to determine the  $\cO(p^4)$
$\chi$PT couplings $L_1$, $L_2$ and $L_3$. The tree-level expression
behaves as $\cO(q^2)$ at large energies, while one-loop corrections can be up to
$\cO(q^{12})$ if spin--1 mesons are included in the theory.
The relevant one-loop topologies are given in Fig.~\ref{fig:scat}.

The study of  the Goldstone scattering is more involved
because the amplitudes depend in this case on two kinematic
invariants, usually chosen among the Mandelstam variables $s$, $t$, $u$.
In order to obtain useful constraints from high-energies, it is convenient to consider
$s\leftrightarrow u$ symmetric amplitudes $T(\nu,t)$,
with $\nu\equiv (s-u)/2$~\cite{RChTb,LECs-Ecker}.
In that case, the forward scattering amplitude $T(\nu,t=0)$ must obey
a once-subtracted dispersion relation
\be
T(\nu,0)\,\, =\,\,T(0,0)\,\, +\,\,  \Frac{\nu^2}{\pi}\Int_0^\infty
\Frac{d\nu^{'2}}{\nu^{'2}}  \,\,\Frac{\mbox{Im}\,T(\nu',0)}{(\nu^{'2}-\nu^2)}\, ,
\ee
with $\nu=s=-u$ for $t=0$.  Thus, at high energies one finds the behaviour
$T(\nu\rightarrow \infty ,0)\sim \nu^0$. Note that
$T(\nu,0)$ can only depend on $\nu^2$ for $s\leftrightarrow u$  symmetric amplitudes.

The scattering amplitude calculated at one-loop within R$\chi$T contains
scalar integrals with up to four propagators. A given one-loop
diagram thus has the general form
\begin{eqnarray}
T^{\rm 1-loop}(s,t,u) &\sim &  a(s,t,u) A_0(M_R)
\nonumber \\
&& \hspace*{-2.5cm}+  b_s(s,t,u) B_0(s)
+ b_t(s,t,u) B_0(t)  +b_u(s,t,u) B_0(u)
\nonumber \\
&& \hspace*{-2.5cm}+  c_s(s,t,u) C_0(0,0,s)+ c_t(s,t,u) C_0(0,0,t)
\nonumber \\
&& \hspace*{-2.5cm}+ c_u(s,t,u) C_0(0,0,u)
\nonumber \\
&& \hspace*{-2.5cm}+  d_{st}(s,t,u) D_0(s,t) + d_{tu}(s,t,u) D_0(t,u)
\nonumber \\
&& \hspace*{-2.5cm}+d_{us}(s,t,u) D_0(u,s)\, ,
\end{eqnarray}
where we have shortened the notation by omitting the mass dependences of the scalar functions
$B_0$, $C_0$ and $D_0$. Likewise, the dependence on the external leg momenta $(p_i^2=0)$
is assumed implicitly. The factors $a(s,t,u)$, $b(s,t,u)$, $c(s,t,u)$ and $d(s,t,u)$
are rational functions with at most resonance double poles
which depend on the structure of the meson vertices.
If we consider the $s\leftrightarrow u$ symmetric amplitude with $t=0$,
the large-$\nu$ expansion of the one-loop scattering amplitude shows the general structure
\begin{eqnarray}
T^{\rm 1-loop}(\nu,0) &=&
\,\, \left(\hat{\kappa}^{(4)}\nu^2+\hat{\kappa}^{(8)} \nu^4 +
\hat{\kappa}^{(12)}\nu^6
 \right)\,\,\,
\ln^2{\Frac{-\nu^2}{M_R^4}}\,
\nonumber \\
&& \hspace*{-1.25cm}
+ \left( \hat{\lambda}^{(4)} \nu^2
+ \hat{\lambda}^{(8)}\nu^4+\hat{\lambda}^{(12)}\nu^6    \right) \,\,\,
\ln{\Frac{-\nu^2}{M_R^4}}\,
\nonumber \\
&& \hspace*{-1.25cm}
+ \left(   \hat{\gamma}^{(4)} \nu^2  + \hat{\gamma}^{(8)} \nu^4
+ \hat{\gamma}^{(12)} \nu^6   \right)
\,\,\,+\cO\left(\nu^0\right)  .
\label{T2}
\end{eqnarray}
It can be shown that for the case of the forward scattering the $\ln^2 (-\nu^2/M_R^4 )$ terms
only arise from the three-point scalar functions.
Local terms from the operators in $\mL^{\mathrm{GB}}$ also contribute to $T(\nu,0)$,
\begin{eqnarray}
\label{op4-scat}
T^{\mathrm{GB}}(\nu,0) &=& \Frac{\widetilde{L}_J\,  \nu^2}{F^4}\, +\,...
\end{eqnarray}
with $\widetilde{L}_J$ the $\cO(p^4)$ coupling or combination of them
for the corresponding scattering amplitude. For instance,
$\widetilde{L}_J=8(2\widetilde{L}_1+2\widetilde{L}_2+\widetilde{L}_3)$ for the
$\pi^0\pi^0\to \pi^0\pi^0$ channel in the chiral limit.

In order to recover the proper short-distance behavior, $T(\nu\rightarrow \infty,0)\sim \nu^0$,
one needs that the bad behaved logarithmic terms get canceled,
which is satisfied by the conditions  $\hat{\lambda}^{(2k)}=\hat{\kappa}^{(2k)}=0$.
The  polynomial pieces can be further made to vanish by establishing relations
among the $\hat{\gamma}^{(2k)}$ coefficients and the couplings $\widetilde{L}_J$
similar to Eqs.~(\ref{LGBfixing}).
Hence, only the well-behaved part of the amplitude, which does not depend on the
$\mathcal{L}^{\mathrm{GB}}$ couplings, determines the scattering amplitude, in agreement with
the requirements of resonance saturation.

More stringent constraints may also be considered for the scattering amplitude.
The absorptive part $\mathrm{Im}\,T$ is given by a sum of positive 
contributions coming from every possible 2-particle cut in the $s$-channel.  
Since there is an infinite number of intermediate states, 
it seems reasonable to expect that the contribution of each of them 
should vanish at high energies 
in order to have a $\cO(\nu^0)$ behaviour for the total absorptive part.

\section{Summary}
\label{sec:summary}

Resonance Chiral Theory supplemented with large-$N_C$ arguments to rule its perturbative expansion
provides a framework to obtain predictions for the low-energy constants of
$\chi$PT in a systematic way. Resonance saturation within this formalism can be defined 
precisely: it states that the $\chi$PT LECs can be written in terms of 
only the resonance couplings and masses. The 
statement is not trivially satisfied because the R$\chi$T  amplitudes
also depend on the parameters $\widetilde{L}_i$, $\widetilde{C}_i,\dots$ of the Goldstone boson
sector which describes the self-interactions of the Goldstone bosons in the presence of resonances.

In this work we have proved that the saturation of the LECs holds at NLO in $1/N_C$, {\it i.e.}
including one-loop corrections in R$\chi$T. Through an analysis
of the analytic structure of the two-point correlators, the pion form-factors and the
Goldstone scattering amplitude we have shown that the values of the parameters
$\widetilde{L}_i$, $\widetilde{C}_i,\dots$ can be fixed once we enforce 
the QCD short-distance behaviour to the corresponding amplitudes.
The resulting R$\chi$T amplitudes get then written only in terms of
parameters related to the dynamics of the resonances, and their
low-energy expansion yields predictions for the LECs.
The matching between R$\chi$T and $\chi$PT is performed at the one-loop
level in both theories, thus ensuring the cancellation of the 
chiral logs. Therefore the NLO estimates for the LECs obtained in this way
do not suffer from the renormalization scale uncertainties inherent to 
the tree-level predictions.

The paper provides further insight on some of the conceptual aspects which have to be addressed in order to devise a consistent 
resonance effective lagrangian at the NLO in $1/N_C$, thus continuing the work initiated in Refs.~\cite{juanjo1,PRD,oneloop}.


\acknowledgments

We want to thank P.~Masjuan, S.~Peris, A.~Pich and J.~Portol\'es for their comments on the manuscript.
This work is supported in part by the Universidad CEU Cardenal Herrera
(grant PRCEU-UCH20/08), by the Generalitat de Catalunya (grant SGR2005-00916),
by the Spanish Government (FPA2007-60323, FPA2008-01430,
the Juan de la Cierva program
and Consolider-Ingenio 2010 CSD2007-00042, CPAN)
and by the EU MRTN-CT-2006-035482 (FLAVIAnet). P.R. acknowledges support from the
DFG Sonder\-forschungsbereich/Transregio~9
``Computergest\"utzte Theoretische Teilchenphysik''.
\vspace*{0.1cm}


\end{document}